\documentclass[a4,twocolumn,pra,aps,showpacs,superscriptaddress,floatfix]{revtex4-1}
\usepackage[utf8]{inputenc}
\usepackage{bm}
\usepackage[usenames]{color}
\usepackage{multirow}
\usepackage{amssymb}
\usepackage{amsbsy}
\usepackage{amsmath}
\usepackage{stmaryrd}
\usepackage{graphicx}
\usepackage{epsfig}
\usepackage{placeins}
\usepackage{ulem}
\usepackage{hyperref}

	 \definecolor{psychedelicpink}{rgb}{0.87, 0.0, 1.0}
	 \definecolor{orange}{rgb}{1,0.5,0}

	\def\be{\begin{equation}}
	\def\ee{\end{equation}}
	\def\bea{\begin{eqnarray}}
	\def\eea{\end{eqnarray}}
	\newcommand{\rl}{\ensuremath{\rangle\langle}}
	\def\Ttr{\textmd{Tr}}

	\def\ttot{\textmd{tot}}
	
	\def\Rre{\textmd{Re}}
	\def\Iim{\textmd{Im}}

	\def\eff{\textmd{eff}}

	\def\ssing{\textmd{sing}}
	
\begin{document}

	\title{Decoherence of an entangled state of a strongly-correlated double quantum dot structure through tunneling processes}

	\author{C. A. B\"usser}
	\email{carlos.busser@gmail.com}
	\author{I. de Vega}
	\author{F. Heidrich-Meisner}
	\affiliation{Department of Physics and Arnold Sommerfeld Center for Theoretical Physics, Ludwig-Maximilians-University Munich, Germany}

	\begin{abstract}
	We consider two quantum dots described by the Anderson-impurity model with one electron per dot. The goal
	of our work is to study the decay of a maximally entangled state between the two electrons localized in the dots. We prepare the 
	system in a perfect singlet and then tunnel-couple one of the dots to leads, which induces  non-equilibrium dynamics. We identify two cases: if the leads are subject to a sufficiently
	large voltage and thus a finite current, then direct tunneling processes cause decoherence and the entanglement as well as spin correlations decay exponentially
	fast. At zero voltage  or small voltages and beyond the mixed-valence regime, virtual tunneling processes dominate and
	lead to a slower loss of coherence. 
	We analyze this problem by studying the real-time dynamics of the spin correlations and the concurrence using two techniques, namely the time-dependent density matrix renormalization group method and
	a master-equation method. The results from these two approaches are in excellent agreement in the direct-tunneling regime for the case in which the dot is weakly tunnel-coupled to the leads.
	We present a quantitative analysis of the decay rates of the spin correlations and the concurrence as a function of tunneling rate, interaction strength, and voltage.

	\end{abstract}
	\maketitle

	\section{Introduction}

	In the last decades, a great effort has been invested in understanding how to utilize 
	the spin degree of freedom of  confined electrons in condensed matter systems  as a component of a quantum computation device or in spintronics 
	\cite{loss98,bennett96,bennett00,cerletti05}.
	In this regard, one of the main challenges  
	is to be able to construct, control and manipulate entangled states between spins. 
	There exist several theoretical proposals and experimental realizations, for instance,   electrons localized in lithographically designed quantum dots~\cite{burkard99,loss00,blaauboer05,okazaki11,amasha13,keller14}, 
quantum dots defined in graphene~\cite{huard07,fritz13}, on carbon nanotubes~\cite{choi05,makarovski07,busser07},  or on organic molecules~\cite{park02,liang02,scott10,zarea13}.
In addition, the charge and spin transport properties of nano-circuits have received great attention due to  their possible applications in electronics and because of  their intrinsic quantum many-body physics \cite{waugh95}. 
	These include the Kondo effect in its many manifestations (see \cite{glazman88,goldhabergordon98,cronenwett98,sasaki00,kouwenhoven01,potok07,busser11,tosi12,ribeiro14} and references therein).

Various ideas of how to create entangled states in nano-structures have been discussed in the literature \cite{bennett96,bennett00,horodecki09,blattmann14,bernien13}. 
As an example, we mention the  proposal to use a Cooper pair splitter to obtain an  entangled pair of  electrons \cite{recher01,oliver02,recher03,herrmann10}. The experimental realization of such a  splitting mechanism is currently at the center of  great efforts~\cite{hofstetter09,hofstetter11,das12}. 
	In other studies,  the possibility of 
	 generating entangled states of electrons localized in quantum dots through non-equilibrium dynamics has been explored, e.g., via the application of a bias potential~\cite{legel07,legel08,busser13}.

\begin{figure}[t]
	\centerline{\includegraphics[width=6cm]{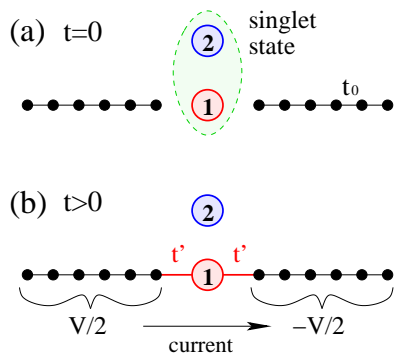}}
	\caption{(Color online) Sketch of the set-up used to study the decoherence process. (a)
The two quantum dots are prepared in a perfect singlet state at $t=0$.
(b) At $t=0^+$, one of the quantum dots is tunnel-coupled  to two leads. The sketch assumes a real-space representation
of the leads and thus this coupling corresponds to adding the tunneling matrix element $t'$ between quantum dot one  and the first site in the left and right lead.
The left and right lead can either be in equilibrium ($V=0$) or be subjected to a finite voltage ($V>0$).
} \label{figure00}
\end{figure}

	Of particular relevance in this context is the analysis of decoherence. This is produced due to the coupling to an environment and may lead to a loss of the information encoded in an entangled state. 
	The purpose of the present work is to analyze the decoherence for the case in which a maximally entangled state 
	between electrons localized in two quantum dots decays because of a tunnel coupling of one of the quantum dots
	to metallic leads. These leads can be either in equilibrium or subject to a voltage.
	In our study, we model the quantum dots using the Anderson-impurity model and we are particularly interested in the
	effect of many-body interactions on the quantum dot. The system is sketched in Fig.~\ref{figure00}.
	Qualitatively, two types of processes that cause decoherence 
	are identified in our study. First, large voltages lead to direct tunneling processes
	accompanied by   a finite electronic current. The decoherence process takes place exponentially fast. For small or vanishing voltages, co-tunneling processes dominate and lead to a slower loss of coherence. For zero voltage and in the Kondo regime,
there is only a partial decay of the entangled state on the time scales studied here.

Our analysis is based on calculating mainly two quantities, first, spin correlations between the two quantum dots and second, the 
concurrence as a measure of entanglement.
We employ two methods in our study, the time-dependent density matrix renormalization group (DMRG) method \cite{daley04,white04,vidal04}
and a weak-coupling master-equation approach (ME). DMRG treats interactions exactly yet is restricted to systems of finite size and thus
it is difficult to access exponentially long times.
We provide a comparison between these two methods and find an excellent agreement for large voltages. At small voltages,
the master equation does not  account for higher order processes in the tunnel coupling between the quantum dot and the environment,
yet the qualitative agreement is nonetheless quite convincing.
We derive a simpler rate  equation in the Markov limit that allows us to develop an intuitive picture for the physical processes causing decoherence and loss of entanglement.
For the regime, in which spin correlations decay exponentially fast, we present an extensive analysis of the dependence of decay rates on model parameters and compare to analytical predictions from the rate equation.

The derivation and applications of quantum master equations for nano-structures coupled to electronic environments has been reviewed in Ref.~\cite{knezevic13}.
The analysis with the ME approach allows to paraphrase our results from the perspective of open quantum systems. Indeed, the direct-tunneling regime with 
an exponential decay of the spin correlations corresponds to a Markovian regime, in which the system  irreversibly looses information. On the contrary, a partial decoherence is related to a strong non-Markovian dynamics, in which the Markov approximation, and even the more accurate weak-coupling ME used here start to fail. 

The formation of entanglement between spins localized in double quantum dots has been studied in a variety of examples by considering a bosonic environment (see, e.g., \cite{contreras08,orth10,fanchini10}). In more detail, \cite{contreras08} analyses the emergence of steady-state entanglement when considering sufficiently strong system-environment couplings, by assuming that the environment always remains in thermal equilibrium and that there are no system-environment correlations. In addition, \cite{orth10} deals with a similar situation and discusses both ground-state and dynamical properties with the  numerical renormalization group (NRG) technique \cite{bulla08}. In the latter case, the (time-dependent) NRG results are compared to the ones of a Redfield equation, which corresponds to the Markov ME that we will derive for our problem. 
In this context, and similar to our case, both methods are found to coincide in the weak-coupling regime, while differences
appear at stronger couplings. However, as we will show in our work, a more general  ME that does not assume Markovian dynamics allows us to reproduce more accurately the DMRG results than the Markovian ME, particularly, around the parameter region where the voltage difference between the leads is similar or equal to the on-site repulsion.

This paper is structured as follows.
 In Sec.~\ref{model}, we introduce the model, definitions, and observables 
and we specify the initial conditions. Section~\ref{sec:dmrg} summarizes technical aspects of our DMRG simulations.
In Sec.~\ref{sec:master}, we derive the master equation
and a simpler rate equation valid in the Markov limit.
Section~\ref{sec:results} summarizes our main results. We first develop a qualitative picture for the decoherence
processes based on the results from the master equation and then present DMRG results for the time dependence of 
spin correlations,  the concurrence, and the electronic current.
These are then compared to numerical solutions of the master equation and we  analyze the decay rates. 
We conclude by a qualitative discussion of the zero-voltage regime. 
Our main results and conclusions are summarized in Sec.~\ref{conclusions}.

\section{Model and definitions}
\label{model}

\subsection{Hamiltonian}

In this work we consider a system of two quantum dots (QD), $i=1,2$, referred to as QD1 and QD2.
The system is schematically  presented in Fig.~\ref{figure00}.
To represent the QDs we use the Anderson-impurity model with an onsite repulsion $U$ (identical for both dots)
and a gate voltage $V_g =- U/2$ such that both dots are at half filling.
 Quantum dot QD1 is connected to the reservoirs via a hybridization term $H_{\rm hy}$ while the leads are described by $H_{\rm B}$
in a tight-binding approximation. 
The total Hamiltonian reads
\begin{eqnarray}
H &=& H_{\rm dots} + H_{\rm B} + H_{\rm hy}  \label{Hamiltonian}\\
H_{\rm dots} &=& \sum_{i=1,2;\sigma} ( V_g n_{i\sigma} +  \frac{U}{2} ~n_{i\sigma} n_{i\bar\sigma } ) \\
H_{\rm B} &=& -t_0 \sum_{\alpha=L,R;\sigma} \sum_{j=1}^N \lbrack(c_{\alpha j \sigma}^\dagger c_{\alpha j+1 \sigma} + \mbox{h.c.})\cr 
&& \quad \quad+ V_{\alpha} n_{\alpha j \sigma}\rbrack \label{Hleads} \\
H_{\rm hy} &=&- t' \sum_{\alpha=L,R;\sigma} ( d^\dagger_{1 \sigma} c_{\alpha 1 \sigma} + {\rm h.c.} )\,. \, \label{VI}
\end{eqnarray}
 $d^\dagger_{i\sigma}$ creates an electron at dot $i$ with spin $\sigma=\uparrow,\downarrow$ and $c^\dagger_{\alpha j\sigma}$ creates an electron with spin $\sigma$ at the site $j$ of the lead $\alpha=L,R$. 
The operator $n_{i\sigma} = d^\dagger_{i\sigma} d_{i\sigma}$ measures the  number of electrons with spin  $\sigma$  on
 dot $i$. The leads and the hybridization are here formulated in real space, and thus $t_0$ and $t'$ are hopping matrix elements 
in the leads and between the first site of each lead and QD1, respectively. $N$ is the number of sites in each lead and the leads 
are half-filled. In DMRG
simulations the leads are taken as finite, while for the master-equation approach we will switch to a momentum-space representation of the leads with a dispersion $\epsilon_{k}=-2t_0 \cos(k)$ (identical for both leads) and the 
leads will be taken as semi-infinite.
The leads can further be subject to a finite
voltage difference $V= V_L-V_R$ with $V_L=-V_R= V/2$. 

For the Anderson-impurity model, it is standard to introduce $\Gamma$, which is the tunneling rate and also measures
the broadening of the dots' levels at the Fermi energy $E_F$ due to hybridization, given by 
\begin{equation}
\Gamma = 2 \pi t'^2 D(E_F)\,, \label{broadening2}
\end{equation}
where $D(\omega)$ is the local density of states of the leads on the first site of the two semi-infinite chains
that model our baths. 
For semi-infinite chains and $V=0$, $E_F=0$, and $D(\omega=0) = 1/\pi t_0$ and thus $\Gamma=2t'^2/t_0$. 
Hereafter, we set $t_0$ and $\hbar$ to unity and measure all quantities in units of $t_0$. Time $t$ is measured
in units of $1/t_0$.
The lattice spacing in the leads is set to unity as well.

\subsection{Initial condition and observables}

The initial condition that we are interested in is a perfect singlet between the electrons localized in dots one and two, 
while the leads are in equilibrium and at zero temperature. Due to this condition, we can view the two sybsystems (leads and dots) as being fully decoupled
at times $t\leq 0$, with no charge fluctuations on the dots.
We then drive the system out-of-equilibrium by coupling QD1 to the leads, which can either be at $V>0$ or $V=0$.

The principal quantity studied in this work is the spin correlation between the dots given by
\begin{equation}
\mbox{S}_{12}(t) = \langle \Psi(t) | \vec{S}_1 \cdot \vec{S}_2  |\Psi(t) \rangle \,.
\end{equation}
$\vec{S}_i$ is the spin-1/2 operator for dot $i$.
In terms of fermionic creation and annihilation operators, $\vec{S}_1 \cdot \vec{S}_2$ can be rewritten as:
\bea
\vec{S}_1 \cdot \vec{S}_2=S^z_1 S^z_2+\frac{1}{2}[S^+_1 S^-_2 + \mbox{h.c.}]
\label{spinspin}
\eea
where we have introduced the usual spin-lowering and raising operators 
 $S_i^+= d^\dagger_{i\uparrow}d_{i\downarrow}$, $S_i^-= d^\dagger_{i\downarrow}d_{i\uparrow}$  and $S^z_i = (n_{i\uparrow}-n_{i\downarrow})/2$.

A  common example for a maximally  entangled state is precisely the singlet of two spin-$1/2$ entities, given by
\bea
|\psi_\ssing\rangle&=&\frac{1}{\sqrt{2}}\left \lbrack |\uparrow\rangle_1|\downarrow\rangle_2-|\downarrow\rangle_1 |\uparrow\rangle_2 \right\rbrack \,.
\label{singlete}
\eea
In order to quantify the entanglement between the QDs, which is directly linked to the spin correlations, 
we use the concurrence \cite{hill97,wootters98,kessler13}. First, we define the single-fermion operator $N_i^s$ on the dots $i=1,2$ as
\begin{equation}
N^s_i = n_{i \uparrow} + n_{i \downarrow} - 2 ~n_{i \uparrow}  n_{i \downarrow}.
\end{equation}
This operator projects onto the subspace with exactly one fermion on each dot. 
Using $N^s_i$, the concurrence can be written as \cite{kessler13}
\begin{equation}
C_{12}(t) = \mbox{max}\left\{0,-\frac{1}{2} - 2 \frac{S_{12}(t)}{\langle \Psi(t)| N^{s}_{1}~N^{s}_{2} |\Psi(t) \rangle} \right\}. \label{concurrence}
\end{equation}
Note that the concurrence takes its maximum value $C_{12}=1$ if ${S_{12}(t)}/{\langle \Psi(t)| N^{s}_{1}~N^{s}_{2} |\Psi(t) \rangle} \to -3/4$. One situation where this result is obtained is when $S_{12} =-3/4$ and $\langle \Psi(t)| N^{s}_{1}~N^{s}_{2} |\Psi(t) \rangle = 1$, such that the spins are in a perfect singlet state, without any charge fluctuations.

We will present results for the current defined as $J=(J_{L,d}+J_{d,R})/2$, i.e., as the average over local currents $J_{L,d}$ and $J_{d,R}$ on the first link in the left and right lead, respectively \cite{alhassanieh06,hm09b},
where these two currents are given by
\begin{eqnarray}
J_{L,d}(t) &=&  i t' \sum_\sigma \langle \Psi(t) | c^\dagger_{L 1\sigma}d_{1\sigma}- d^\dagger_{1\sigma}c_{L 1\sigma}  | \Psi(t) \rangle \\
J_{d,R}(t) &=&  i t' \sum_\sigma \langle \Psi(t) |d^\dagger_{1\sigma}c_{R 1\sigma} - c^\dagger_{R 1\sigma}d_{1\sigma} | \Psi(t) \rangle.
\label{current}
\end{eqnarray}

\section{Density Matrix Renormalization Group}
\label{sec:dmrg}

We use time-dependent DMRG \cite{vidal04,daley04,white04} to obtain the steady state in the presence of a finite bias voltage. 
Standard DMRG  is a numerical technique designed to calculate the ground state of strongly-correlated systems by efficiently
representing wave functions in a truncated but optimized basis \cite{white92,schollwoeck05}. This effectively
uses matrix-product states as the underlying ansatz wave functions \cite{schollwoeck11}. 
A particular advantage of DMRG is that the accuracy of how wave functions and observables are approximated
can be controlled and in principle be made arbitrarily small by tuning the so-called discarded weight \cite{schollwoeck05}.

In a first step, we use DMRG to calculate the ground state $|\Psi_0\rangle$ of the Hamiltonian  Eq.~\eqref{Hamiltonian} 
with $t'=0$ in the presence of an auxiliary term $H_J= J_0 \vec{S}_1 \cdot \vec{S}_2$.
Choosing a large $J_0\gg U$,
we prepare the initial singlet state with  $\langle \vec{S}_1 \cdot \vec{S}_2 \rangle= -3/4$.
At time $t=0$, a quench in the Hamiltonian is performed: the coupling $J_0$ is removed and QD1 is connected to the leads, i.e., $t'$ is set to a finite value.

The time evolution 
\begin{equation}
|\Psi(t)\rangle = \mbox{e}^{-iHt} |\Psi_0\rangle 
\end{equation}
is performed
using a Trotter-Suzuki breakup of the time-evolution operator \cite{vidal04}.
In order to implement the Trotter-Suzuki expansion,  the two  QDs are treated as one site, i.e.,
our one-dimensional system consists of the $2N$ sites in the leads with four local states and one site representing the dots 
with eight local states.

We keep 500 states for the calculation of the ground state and a maximum of 2000 states for the time evolution. 
We verified that the truncation error is at least below $10^{-5}$ during the DMRG calculations, which we have found to be  sufficient to ensure 
reliable numerical results.
The time step in the  DMRG time evolution  was set to $\delta t \sim 0.1$, which is much smaller than any relaxation time found in the  present problem. 

The DMRG method has been successfully  used to study non-equilibrium transport through nano-structures with electronic correlations \cite{alhassanieh06,hm09,dasilva08,kirino08,boulat08,hm10,branschaedel10,einhellinger12,nuss13,canovi13,hm09b,eckel10}. 
These applications include the calculation of current-voltage characteristics for the interacting resonant-level model \cite{boulat08}, the single-impurity Anderson
model \cite{kirino08,hm09b,einhellinger12,nuss13} as well as multiple-dot systems \cite{busser13,canovi13,hm10}. 
A comparison with various other numerical methods \cite{eckel10} shows that DMRG reliably captures the steady-state currents
for voltages larger than Kondo temperature and in the mixed-valence regime of the single-impurity Anderson model.
The Kondo regime can in principle be accessed with tDMRG by using so-called Wilson leads \cite{dasilva08}.
Alternatively, one may resort to the time-dependent numerical renormalization group method, which has been used to study
relaxation dynamics in the Kondo regime \cite{anders05,anders06}.

\section{Master equation formalism}\label{ME}
\label{sec:master}
The dynamics of  molecules and quantum dots interacting with a spin environment or two metal leads has been analysed in many previous studies \cite{nitzan2001,pecchia2004,ratner2005,stanek2013}. Some of them are based on considering the quantum dots as an open quantum system (OQS) coupled to a reservoir of electrons \cite{merkulov2002,khaetskii2003,loss2004,fischer2008,faribault2013,hackmann2014}.  Then, the dynamics of the quantum dots can be analysed using a quantum master equation  \cite{gurvitz1996,pedersen2005,li2005}, that evolve their reduced density operator. Due to its relatively simple structure, the ME provides an intuitive understanding of the system dynamics. Of special interest to our discussion are those analyses that describe non-Markovian effects arising from the finite environment memory time \cite{breuerbook,rivas2011a,breuer2012}. These occur because the electronic environment may not recover instantaneously from the coupling with the system, so that, contrary to what occurs due to Markovian interactions described with a Lindblad formalism \cite{lindblad1976,breuerbook}, some back-flow of information may occur from the environment to the system \cite{breuer2009,rivas2010}. Among these types of analyses including non-Markovian effects, different master equations have been proposed to determine the evolution of the reduced density matrix of the system, in our case the quantum dots. Some models are specific to a spin-star configuration, where a single spin-less quantum dot (i.e., a two level system or spin-$1/2$) is considered to be coupled to a fermionic environment. For instance,  in \cite{breuer2004} it is shown how the dynamics of the central spin-$1/2$ coupled to the spin environment through a Heisenberg XX interaction can be solved exactly, and the result is compared to the one obtained from a perturbative formulation. In \cite{harbola2006} two  quantum dots with spinless fermions linearly coupled to an environment with a more general coupling Hamiltonan than the Heisenberg XX interaction are analyzed. This description is based on approximating the total state as a product state of the form $\rho_\ttot(t)=\rho_{\rm B}\otimes \rho_{\rm S}(t)$. Here, the quantity $\rho_{\rm B}$ represents the environment state, considered to be always in  equilibrium, and an evolution equation is derived for $\rho_{\rm S}$, which represents the reduced density matrix of the OQS.  An alternative derivation is found in \cite{jinshuang2008}, where the master equation is written in terms of  an operator that can be computed based on  a set of hierarchically coupled equations. Although this formulation leads to an exact and numerically tractable way to deal with the reduced density matrix dynamics, it is particularly suitable  for system-environment couplings that can be described with an exponential correlation function \cite{haertle2013}. 

In this section, we first derive a master equation in the weak-coupling limit $\Gamma \ll U$. Then, we discuss
simplifications that allows us to arrive at a rate equation.

\subsection{Derivation}
Let us divide  the Hamiltonian Eq.~(\ref{Hamiltonian})  into two different contributions, $H=H_0+H_{\rm p}$, where the unperturbed part $H_0=H_{\rm dots} + H_{\rm B}$ represents the sum of the Hamiltonian of the dots and the Hamiltonian of the leads, and $H_{\rm p}=H_{\rm hy}$ represents the perturbation
that couples leads and QD1.
Then, the leads can be considered as an environment coupled to an open quantum system, 
in this case QD1 and QD2. In order to describe this problem with the theory of open quantum systems 
we diagonalize the Hamiltonian of the leads by going to a momentum-space representation using 
$c_{\alpha j \sigma}=(1/\sqrt{N})\sum_k e^{ik j}c_{\alpha k \sigma}$.  $N$ thus equals  the number of modes in the environment. 

In this basis,  the Hamiltonian from Eq.~\eqref{Hamiltonian} can be written as $H=H_0+H_{\rm hy}$ with 
\bea
H_0&=&H_{\rm dots}+H_{\rm B}\nonumber\\
H_{\rm hy}&=&-t' \sum_k \sum_{\alpha=L,R;~\sigma} ( d^\dagger_{1 \sigma} c_{\alpha k \sigma} +{\rm h.c.} ) 
\label{hamilME}
\eea
where $H_{\rm B}=\sum_{\alpha k \sigma} \epsilon_{k} c^\dagger_{\alpha k\sigma}c_{ \alpha k\sigma} $.

We consider an initial state of the form 
\bea
\rho(0)=\rho_{\rm B}(0)\otimes\rho_{\rm dots}(0),
\label{init}
\eea
expressed with density matrices $\rho$ for the full system, $\rho_{\rm B}$ for the baths, and $\rho_{\rm dots}$ for the quantum dots.
Here, $\rho_{\rm B}(0)=\rho_{L}(0)+\rho_{R}(0)$ describing the initial state of the leads $\rho_{\alpha}(0)$, with $\alpha=L,R$ at zero temperature, and $\rho_{\rm dots}(0)=|\psi_\ssing\rangle\langle \psi_\ssing|$, with the singlet defined in Eq.~(\ref{singlete}), such that 
\bea
\rho_{\rm dots}(0)&=& |\psi_\ssing\rangle\langle\psi_\ssing| \cr
 &=&  \frac{1}{2} \left\lbrack\rho^{(1),1}\otimes\rho^{(2),4} -\rho^{(1),2}\otimes\rho^{(2),3} \right. \cr
&& \left.- \rho^{(1),3}\otimes\rho^{(2),2}+\rho^{(1),4}\otimes\rho^{(2),1}
\right\rbrack\,.
\label{singletM}
\eea
Here, we have defined the projectors $\rho^{(i),n}$
\begin{eqnarray}
\rho^{(i),1} &=& | \uparrow\rangle_i \langle \uparrow | \cr
\rho^{(i),2} &=& | \uparrow\rangle_i \langle \downarrow | \cr
\rho^{(i),3} &=& | \downarrow\rangle_i \langle \uparrow | \cr
\rho^{(i),4} &=& | \downarrow\rangle_i \langle \downarrow |\label{eq:four}\,. 
\end{eqnarray}
Assuming that the tunnel matrix element $t'$ between the environment, i.e., the leads, and QD1 is very small 
as compared to $t_0$ such that typical time-scales induced by 
$t'$ are much slower  than time scales of the bath,
one may obtain a closed evolution for the reduced density matrix of the quantum dots $\rho_{\rm dots}=\Ttr[\rho]$. 
 This equation, to second order in $t'$ and thus first order in $\Gamma$, can be written as 
\begin{eqnarray}
&&\frac{d\rho_{\rm dots}(t)}{dt}= \cr
&&-  \int_{0}^{t}d\tau\, \mbox{Tr}_{\rm B}\{[{H}_{\rm hy} (t),[{H}_{\rm hy} (t-\tau),\rho_{\rm B}(t)\otimes\rho_{\rm dots} (t)]]\},\cr
&&\label{total5}
\end{eqnarray}
Here, we have taken the  Born approximation,
where the environment is considered to be unperturbed by the interaction with the system, and therefore, 
\bea
\rho(t)=\rho_{\rm B} \otimes \rho_{\rm dots}(t)\,.
\label{born}
\eea
 $\rho_{\rm B}=\rho_{\rm B}(0)$ is assumed to always describe an equilibrium state. 

Here,  $ H_{\rm hy}(t)$ is expressed  in the interaction picture with respect to $H_0$
\bea
 H_{\rm hy}(t) =-t' \sum_k\sum_{\alpha=L,R;~\sigma} ( d^\dagger_{1 \sigma}(t) c_{\alpha k \sigma}(t) +{\rm h.c.} )\,,
\eea
with $A(t)=e^{iH_0 t}A e^{-iH_0 t}$, where $A$ is any observable  operating on the system or on the environment Hilbert space. 
This expression is inserted into  Eq.~(\ref{total5}) and we consider also that 
\bea
&&\mbox{Tr}_{\rm B}\{[ {H}_{\rm hy} (t),[{H}_{\rm hy} (s),\rho_{\rm B}\otimes\rho_{\rm dots} (t)]]\}\cr
&=& \mbox{Tr}_{\rm B}\{ H_{\rm hy} (t)H_{\rm hy} (s) \rho_{\rm dots}(t)\rho_{\rm B}
-H_{\rm hy} (t)\rho_{\rm dots}(t)\rho_{\rm B} H_{\rm hy}(s) \cr
&-&H_{\rm hy} (s)\rho_{\rm dots}(t)\rho_{\rm B} H_{\rm hy}(t)
+\rho_{\rm dots}(t)\rho_{\rm B}H_{\rm hy}(s)H_{\rm hy}(t) \}, \nonumber
\eea
where we have defined $s=t-\tau$, and  
\bea
&&\mbox{Tr}_{\rm B}\{\rho_{\rm B} c_{\alpha k\sigma}c_{\alpha' k'\sigma'}\}=0;\cr
&&\mbox{Tr}_{\rm B}\{\rho_{\rm B} c^\dagger_{\alpha k\sigma}c_{\alpha' k'\sigma'}\}=\delta_{k,k'}\delta_{\alpha,\alpha'}\delta_{\sigma\sigma'}f_\alpha(\epsilon_k),
\eea
where $f_\alpha(\epsilon_{k})=[\exp({\beta}(\epsilon_{k}-V_\alpha))+1]^{-1}$ is the  Fermi distribution function number for lead $\alpha$, with ${\beta}=1/(k_BT)$, where $k_B$ is the Boltzmann constant and $T$ the environment temperature.

After some standard manipulations, and by going to the Schr\"odinger picture, the master equation can be written as 
\begin{eqnarray}
&&\frac{d\rho_{\rm dots}  (t)}{dt}=-i[H_{\rm dots} ,\rho_{\rm dots} (t) ]\cr 
 &+&\sum_{\alpha,\sigma}\int_0^t d\tau\, G_\alpha^{+*}(t-\tau)
\times [d_{1\sigma}^\dagger ,\rho_{\rm dots} (t)  d_{1\sigma}(\tau-t)]\cr
&+&\sum_{\alpha,\sigma}\int_0^t d\tau\, G_\alpha^{+}(t-\tau) 
\times[d_{1\sigma}^\dagger(\tau-t) \rho_{\rm dots} (t) ,d_{1\sigma}]\cr
&+&\sum_{\alpha,\sigma}\int_0^t d\tau\, G_\alpha^{-}(t-\tau)
\times[d_{1\sigma}(\tau-t)\rho_{\rm dots} (t) ,d_{1\sigma}^{\dagger}] \cr
&+&\sum_{\alpha,\sigma}\int_0^t d\tau\, G_\alpha^{-*} (t-\tau)
\times[d_{1\sigma},\rho_{\rm dots} (t) d_{1\sigma}^{\dagger}(\tau-t)]\cr &+&{\mathcal O}(t'^4),
\label{master1}
\end{eqnarray}
with 
\begin{eqnarray}
G_\alpha^- (t-\tau)=t'^2\sum_k 
(1-f_\alpha(\epsilon_k))e^{-i\epsilon_k (t-\tau)},
\label{correlplus}
\end{eqnarray}
and 
\begin{eqnarray}
G_\alpha^+ (t-\tau)=t'^2\sum_k f_\alpha(\epsilon_{k})e^{i\epsilon_k (t-\tau)},
\label{correlminus}
\end{eqnarray}
where now we have not assumed the interaction picture with respect to the system. 
In order to numerically compute these quantities, we take the thermodynamic limit (large $N$) and 
replace sums by integrals.
In our work, we consider the case at $T=0$ where the number of quanta in the mode $\epsilon_k$ is only different from zero when the frequency is below the bias potential corresponding to the lead $\alpha$, i.e., $f_\alpha(\epsilon_k)=\theta(\epsilon_k-V_\alpha)$.

We note that this master equation is identical to the one corresponding to a bosonic environment \cite{yu1999}, 
by replacing the Fermi-function by a Bose-distribution function.

\subsection{Markov limit and rate equations}
\label{sec:rate_eq}
We now consider a mean-field approximation for the interaction term in $H_{\rm dots}$, i.e.,
\begin{eqnarray}
\frac{U}{2}\sum_{i,\sigma}n_{i\sigma} n_{i\bar{\sigma}}  
&\approx&  \frac{U}{2}\sum_{i,\sigma}(n_{i\sigma} \langle n_{i\bar {\sigma}}\rangle+\langle n_{i\sigma}\rangle n_{i\bar{\sigma}}) \cr 
 && - \frac{U}{2} \sum_{i,\sigma} \langle n_{i{\sigma}} \rangle \langle n_{i\bar {\sigma}}\rangle\cr 
&=&U\sum_{i,\sigma}n_{i\sigma} \langle n_{i\bar {\sigma}}\rangle  \cr
&& - \frac{U}{2} \sum_{i,\sigma} \langle n_{i {\sigma}} \rangle \langle n_{i\bar {\sigma}}\rangle\,.
\end{eqnarray} 
Within this approximation, the time evolution of the creation and annihilation operators acting on QD1 
with respect to $H_{\rm dots}$ can be written as: 
\bea
d_{1\downarrow} (t)&=&e^{-i(V_g+{U}\langle n_{i\uparrow}\rangle)t}d_{1\downarrow}\cr
d_{1\uparrow} (t)&=&e^{-i(V_g+{U}\langle n_{i\downarrow}\rangle)t}d_{1\uparrow}\,.
\eea
For the sake of clarity, we have reinserted the gate voltage $V_g$ although in all numerical simulations, we set $V_g=-U/2$. 
Considering that for $V_g=-U/2$,  
$\langle n_{i\sigma}\rangle=1$ and $\langle n_{i\bar\sigma}\rangle=0$,  
this simplifies to $d_{1\sigma} (t)=e^{-i\Omega_\sigma t}d_{1\sigma}$, 
with the effective single-particle  levels $\Omega_{\sigma}$ on the dot
\begin{equation}
\Omega_{\sigma}=V_g;\quad \Omega_{\bar \sigma} = V_g +U \,.
\end{equation}
 Note also that this mean-field approximation is only valid when the fluctuations in the particle number on 
each quantum dot are negligible. Since the system's Hamiltonian $H_{\rm dots}$ 
is diagonal in $n_{i\sigma}$ and hence 
does not produce any  such fluctuations, these can only  originate from the hybridization with the environment. 
However, if the environment is sufficiently Markovian as it happens for $V>U$ (to be verified later by comparison with DMRG), 
it will just induce dissipation and decoherence in the system, and therefore  
will not be able to create any coherence in its observables. Hence, being in the Markovian limit or in a limit where the weak-coupling approximation is valid, also assures us that the  mean-field approximation is a reasonable approach. 

Using this approximation, the master equation Eq.~(\ref{master1}) can be rewritten as
\begin{eqnarray}
\frac{d\rho_{\rm dots}  (t)}{dt}&=&-i[H_{\rm dots} ,\rho_{\rm dots} (t) ]\cr 
&+&\sum_{\alpha,\sigma}\gamma^{+*}_{\alpha\sigma}(t) \times [d_{1\sigma}^\dagger ,\rho_{\rm dots} (t)  d_{1\sigma}] \cr 
&+&\sum_{\alpha,\sigma}\gamma^{+}_{\alpha\sigma}(t) \times [d_{1\sigma}^\dagger \rho_{\rm dots} (t) ,d_{1\sigma}] \cr
&+&\sum_{\alpha,\sigma}\gamma^{-}_{\alpha\sigma}(t) \times [d_{1\sigma}\rho_{\rm dots} (t) ,d_{1\sigma}^{\dagger}] \cr
&+&\sum_{\alpha,\sigma}\gamma^{-*}_{\alpha\sigma}(t) \times [d_{1\sigma},\rho_{\rm dots} (t) d_{1\sigma}^{\dagger}] \cr 
&+& {\mathcal O}(t'^4),
\label{master2}
\end{eqnarray}
where 
\bea
\gamma^{+}_{\alpha\sigma}(t)=\int_0^t d\tau\, G_\alpha^{+}(t-\tau) e^{-i\Omega_{\sigma} \tau};\cr
\gamma^{-}_{\alpha\sigma}(t)=\int_0^t d\tau\, G_\alpha^{-}(t-\tau) e^{i\Omega_{\sigma} \tau}\,.
\label{rates1}
\eea
As discussed in Appendix~\ref{sec:markov}, this master equation can be further simplified when going deeper in the Markov regime, where the decay
 of the functions $G_\alpha^\pm(t-\tau)$, i.e., the correlation time of the environment, is negligible 
compared to  the time scales of variations given by the system's eigenenergies $\Omega_\sigma$.

The density matrix $\rho_{\rm dots}(t)$ is in principle defined in the Hilbert space of the two quantum dots. 
However, we note that while the initial state Eq.~(\ref{init}) corresponds to an entangled state, only QD1 is coupled to the leads. Considering this fact, the time evolution of the initial state Eq.~(\ref{init}) will only affect each of the first terms of 
Eq.~(\ref{singletM}) corresponding to QD1. 
Hence, according to Eq.~(\ref{singletM}) and due to the Born approximation Eq.~(\ref{born}), 
the full density matrix  $\rho(t)=\rho_{\rm B}(0)\otimes\rho_{\rm dots}(t)$ can be reexpressed using
\bea
\rho_{\rm dots}(t)&=&\frac{1}{2}\bigg[\rho^{(1),1}(t)\otimes\rho^{(2),4}-\rho^{(1),2}(t)\otimes\rho^{(2),3} \cr
 &&-\rho^{(1),3}(t)\otimes\rho^{(2),2}+\rho^{(1),4}(t)\otimes\rho^{(2),1}\bigg]
\label{rhostot}
\eea
where $\rho^{(1),n}(t)$  are the time-propagated  four density matrices defined in the Hilbert space of QD1 
with the initial conditions specified in Eq.~\eqref{eq:four}.
The $\rho^{(1),n}(t)$ are evolved according to the master equation Eq.~(\ref{master1}).
In the former expression, the $\rho^{(2),n}$ do not evolve such  that each matrix $\rho^{(2),n}$ is still given by Eq.~(\ref{eq:four}). 
The density matrix Eq.~(\ref{rhostot}) can be used to calculate the expectation values 
of any operator acting on the two quantum dots and, in particular, the spin-correlation operator defined in Eq.~(\ref{spinspin}). 
The spin-spin correlator can be expressed in terms of matrix-elements $\rho^{(1),n}_{\sigma\sigma'}$  of the $\rho^{(1),n}(t)$ 
\bea
S_{12}(t)&=&\frac{1}{4}\bigg[(\rho^{(1),1}_{\uparrow\uparrow}-\rho^{(1),4}_{\uparrow\uparrow})+(\rho^{(1),4}_{\downarrow\downarrow}-\rho^{(1),1}_{\downarrow\downarrow})\bigg] \cr
&&+\Rre\left[\rho_{\uparrow\downarrow}^{(1),2} \right]\,.
\eea
defined above and evolved in the Hilbert space of QD1. 
As noted above, in our problem, there is no finite spin polarization on the dots and the associated $U(1)$ symmetry is not broken. This 
symmetry is also preserved in the mean-field approximation employed to derive the rate equation, and therefore,
$\rho^{(1),1}_{\uparrow\uparrow}=\rho^{(1),1}_{\downarrow\downarrow}$ and $\rho^{(1),4}_{\uparrow\uparrow}=\rho^{(1),4}_{\downarrow\downarrow}$, 
such that we finally obtain 
\begin{equation}
S_{12}(t)=\Rre\left[\rho_{\uparrow\downarrow}^{(1),2}\right]\,.
\label{eq:rate_me}
\end{equation}
In the Markov limit that is discussed in detail in Appendix~\ref{sec:markov}, we demonstrate that
$\rho_{\uparrow\downarrow}^{(1),2}$ decays exponentially with time since 
\begin{equation}
\frac{d \rho_{\uparrow\downarrow}^{(1),2}}{dt} =  -\sum_{\alpha,\sigma}
({\Gamma}^-_{\alpha\sigma}+{\Gamma}^+_{\alpha\sigma})\rho_{\uparrow\downarrow}^{(1),2}
\label{eq:rate}
\end{equation}
and the rates $\Gamma_{\alpha\sigma}^{\pm}$ take a simple expression
\bea
&& \Gamma^-_{\alpha\sigma}=t'^2D(\Omega_{\sigma})(1-f_\alpha(\Omega_{\sigma}))\cr
&& \Gamma^+_{\alpha\sigma}=t'^2D(\Omega_{\sigma})f_\alpha(\Omega_{\sigma})\,.
\label{rates3}
\eea
The expression for $D(\omega)$ is 
\bea
D(\omega)=\Iim\bigg[\frac{\omega-\sqrt{\omega^2-4t_0^2}}{2\pi t_0^2}\bigg]\,.
\label{DOS}
\eea
As a result, we obtain 
\begin{equation}
S_{12}(t) = S_{12}{(t=0)}\, \mbox{exp}(-\gamma_0 t)
\label{eq:s12fit}
\end{equation}
where the rate $\gamma_0$ is given by
\begin{equation}
\gamma_0 = \sum_{\alpha, \sigma} (\Gamma_{\alpha\sigma}^{+} + \Gamma_{\alpha\sigma}^{-})\,.
\end{equation}

\begin{figure}[t]
\centerline{\includegraphics[width=8.5cm]{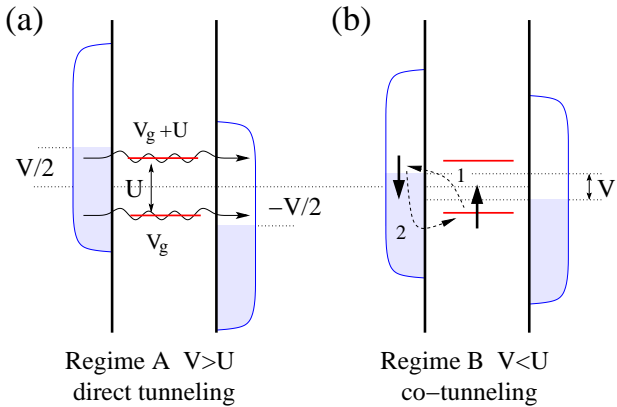}}
\caption{ (Color online)
Sketch of the two different regimes for the decoherence process.
(a) In regime A ($V>U$), the decoherence and loss of entanglement is due to a finite current and thus direct tunneling processes.
(b) In  regime B, realized for $0\leq V<U$ and $\Gamma \ll U$,
virtual tunneling processes involving the many-body interaction $U$ cause the loss of decoherence. An example of such a co-tunneling process is shown in (b), in which an electron with spin up tunnels out of the dot (arrow 1) and  one with spin down tunnels in (arrow 2), mediating a spin-flip process.
} \label{figure01}
\end{figure}

Upon inspection of Eq.~\eqref{rates3}, we realize that for a symmetrically applied voltage ($V_L=V_R=\pm V/2$),
for one lead, at most  one of the rates $\Gamma_{\alpha\sigma}^{\pm}$ can be non-zero, as depicted in Fig.~\ref{figure01}.
This happens for $V>U$,  while  for $0\leq V<U$, both rates vanish.
In the following we will refer to these  qualitatively different regimes as regime A ($V>U$) 
and regime B ($V<U$) [these regimes will be discussed in more detail in Sec.~\ref{sec:regimes}]. Therefore,
the rate equation Eq.~\eqref{eq:rate}  only accounts for direct tunneling with no renormalization of the dot's levels, resulting
in an infinite life-time of spin correlations in regime B and and an exponential decay with a rate $\gamma_0$ in regime A.
Overall, since each lead by symmetry contributes equally to the total rate (in other words, $D(V_g) = D(V_g+U)$), we find for regime A:
\begin{equation}
\gamma_0 = 2\Gamma D(V_g)\,. 
\label{eq:rate-regA}
\end{equation}
Regimes A and B will be further discussed in the next section.

\section{Numerical results}
\label{sec:results}
In this section we present our numerical results for the time dependence of spin correlations,
the concurrence, and the current. We obtain these data from DMRG simulations, complemented
by numerical solutions of the master equation Eq.~\eqref{master1}. We first provide a qualitative discussion
of the possible parameter regimes and the processes that lead to decoherence in Sec.~\ref{sec:regimes}. Then we present DMRG
data in Sec.~\ref{sec:dmrg_data} and  compare results from both methods in Sec.~\ref{sec:compare}. 
In Sec.~\ref{sec:rates}, we analyze the dependence of the decay rates $\gamma_0$ and $\gamma_c$
of spin correlations and of the concurrence, respectively, on $U$, $\Gamma$, and voltage $V$.
Finally, we present a qualitative discussion of the Kondo regime $V< T_K$, where $T_K$ is the Kondo
temperature.

\begin{figure}[t]
\centerline{\includegraphics[width=0.95\columnwidth]{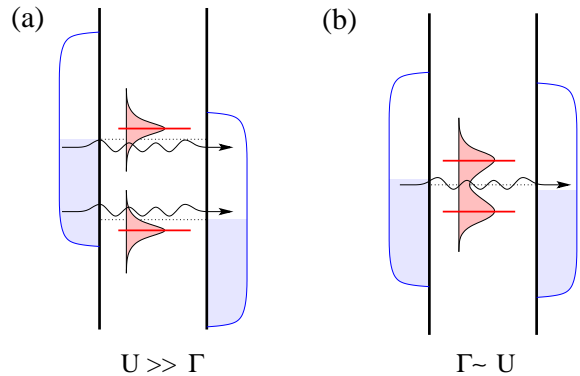}}
\caption{ (Color online)
This figure illustrates the effect of a finite and large $\Gamma$ in regime B ($V<U$).
While for $U/\Gamma \gg 1 $, only co-tunneling processes are possible, a finite $\Gamma$ results
in a broadening of the levels at $\Omega_{\sigma}= V_g$ and $\Omega_{\bar\sigma}=V_g +U$, thus allowing for
tunneling off resonance. Typical situations are sketched for (a) the Kondo regime ($U/\Gamma \gg 1$) and $V\propto U$ 
and (b) the mixed-valence regime $U\sim \Gamma$ and  $0\leq V < U$.
}\label{figure05}
\end{figure}

\subsection{Parameter regimes}
\label{sec:regimes}

Based on the analysis of the master equation in the Markov limit from Sec.~\ref{sec:rate_eq}, 
we expect two different regimes for the loss of entanglement and decay of correlations upon coupling  QD1 to
the reservoirs.
These two regimes are schematically depicted in Fig.~\ref{figure01}. 
In regime A [see Fig.~\ref{figure01}(a)], realized for $V>U$, direct tunneling processes are possible and a finite current will
flow through the  many-body levels $\Omega_{\sigma}=V_g$ and $\Omega_{\bar\sigma}=V_g+U$. 
The decoherence process is controlled by the rate of the electrons hopping on and off the dot, given by $\Gamma$, which 
can flip the spin and will thus destroy the initial singlet.
In regime B  [see Fig.~\ref{figure01}(b)], realized for $0\leq V<U$, 
there cannot be any current as long as we neglect any broadening of QD1's levels due to hybridization.
Therefore, virtual processes will be relevant that can also flip the spin of the electron in QD1.
Such processes are also referred to as co-tunneling, a simple perturbative estimate of characteristic time scales is $U/t'^2 \propto U/\Gamma$. Within the regime of validity of the master equation,  under these assumptions and  in regime B, no current can flow.

A finite $\Gamma>0$ will cause a broadening of the levels. This has to develop dynamically on time scales proportional to $1/\Gamma$ 
since the quantum dots are initially isolated from the environment in our set-up. This broadening will give rise to tunneling
in regime B for voltages $V \lesssim V$. Schematic examples are shown in Fig.~\ref{figure05}.

In the limit of very low voltages, one may also encounter Kondo physics \cite{hewsonbook}. This requires $V<T_K$ and would correspond
to the dynamical formation of the Kondo resonance in QD1's local density of states, allowing for a finite current even for $U\gg \Gamma$.
However, this physics will manifest itself on time scales of $1/T_K$ \cite{anders06,dasilva08} much larger than what can be accessed with DMRG
using tight-binding leads if $U\gg \Gamma$.
Therefore, our DMRG results for  $U\gg \Gamma$ and $V<T_K$ only capture the short-time dynamics correctly.
Kondo physics and co-tunneling are also not captured  in our master-equation approach since these involve higher-order processes and require a resummation.

\begin{figure}[t]
\centerline{\includegraphics[width=7cm]{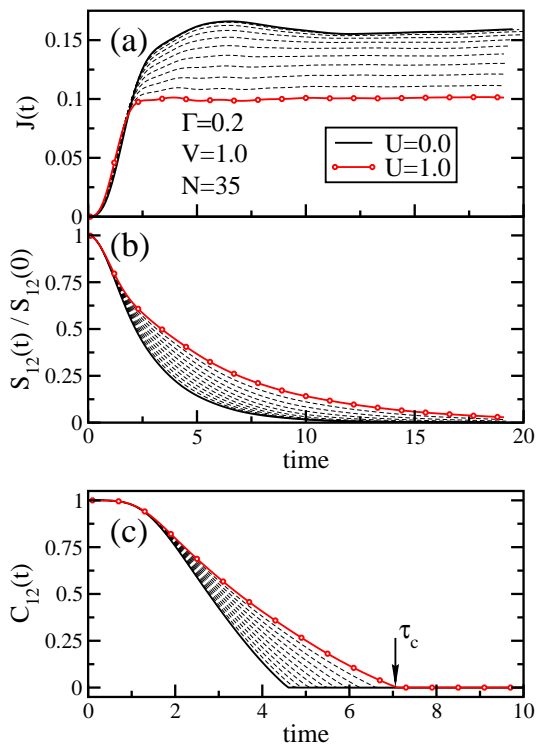}}
\caption{ (Color online)
DMRG data for  (a) the current $J(t)$, (b) spin correlations $S_{12}(t)$ and (c) concurrence $C_{12}(t)$ versus time for several values of $U$ and  $\Gamma=0.2$, $V=1$
for a system with $N=35$ 
($U=0, 0.1, 0.2, \dots ,1$).
These parameters ($V>U$) put the system into regime A in the spirit of Sec.~\ref{sec:regimes} or at its boundary $V=U$.
 The spin correlations are normalized to their value at $t=0$, $S_{12}(t=0)=-3/4$ and decay to zero, which is the faster
the smaller $U$ is.
The concurrence $C_{12}(t)$ drops to zero instantaneously at a time $\tau_c$, which depends on $U$.
} \label{figure02}
\end{figure}

\subsection{DMRG results  for the time evolution of correlations, concurrence, and current}
\label{sec:dmrg_data}
\subsubsection{Time-evolution of correlations, concurrence, and current}
\label{sec:expo}
Figure~\ref{figure02} shows results for the current, the concurrence and the spin correlations calculated for fixed $\Gamma=0.2$ and $V=1$ and considering several values of the Coulomb repulsion $0 \leq U \leq 1$ such that these data are for regime A.
We start by discussing the current displayed in Fig.~\ref{figure02}(a). The current first undergoes transient dynamics
 and then takes a quasi-stationary value (i.e., a plateau in time), which we refer to as the steady-state regime  \cite{alhassanieh06,hm09,hm09b}.
Since the leads have a finite length in these simulations, 
there is a system-size dependent revival time, resulting in a decay of the steady-state
current and a sign change \cite{alhassanieh06,branschaedel10}. 
This effect (realized for $t \gtrsim 35$ for the parameters of the figure) is not shown in Fig.~\ref{figure02}(a).
For a discussion of such transient time scales, as well as a comprehensive analysis of the time-dependent behavior for currents, see Refs.~\cite{hm09} and \cite{branschaedel10} and references therein.

\begin{figure}[t]
\centerline{\includegraphics[width=8cm]{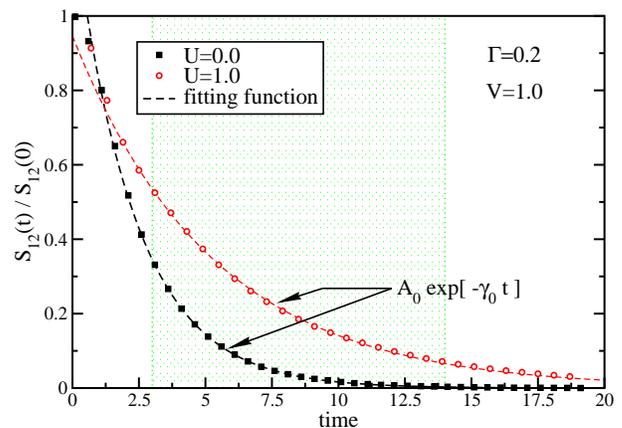}}
\caption{ (Color online)
Spin correlations $S_{12}(t)/S_{12}(0)$ versus time for $U=0$ (squares) and $U=1$ (circles)
at $V=1$ and for $\Gamma=0.2$ (DMRG data). 
The dashed lines are fits to the data using an exponential $f(t) = A_0\, \mbox{exp}(-\gamma_0 t)$. These fits are used to
extract the decoherence rate $\gamma_0$ of the spin correlations from the data.
The time window over which these fits describe the data the best are indicated by the  shaded area.
} \label{figure03}
\end{figure}

While the main purpose of our present work is to understand the time evolution of entanglement properties and spin correlations,
note that we also, as a by-product, obtain the current-voltage characteristics of a single Anderson impurity. Our way of driving
the system out-of-equilibrium is different from other DMRG studies \cite{hm09b,branschaedel10} since
there, typically the quantum dot is connected  to the leads via $t'>0$ in the initial state. Hence, in these other studies,  the initial state is not a product state
between system and environment. The transient dynamics, comparing Ref.~\cite{hm09b,branschaedel10} and our case, are different  in two respects.
 First, the short-time increase of $J \propto t^2$ is quadratic in time in our case,
a direct consequence of the product form of the initial state as compared to $J\propto t$ observed in Refs.~\cite{alhassanieh06,hm09b,kirino08}.
Second, the transients exhibit no 'overshooting' (i.e., $J(t)$ first going well beyond the steady-state value),
 which typically occurs for the initial conditions of Refs.~\cite{hm09b,branschaedel10} (see also Ref.~\cite{nuss13} for a DMRG study 
of different initial conditions).
This makes it easier to extract the steady-state current. 
We have verified that the steady-state current obtained from our initial 
conditions is identical to the results of Ref.~\cite{hm09b,eckel10} for large voltages $V>U$ and in the mixed-valence regime $U\sim \Gamma$.

In Fig.~\ref{figure02}(b), we present the spin correlations for the same parameters as in Fig.~\ref{figure02}(a). 
This quantity also exhibits a short-time dynamics that is independent of $U$. After this first transient, an exponential decay emerges,
as expected from the rate equation Eq.~\eqref{eq:rate}.
 To make this important point more transparent,  we present selected results for the behavior of $S_{12}(t)$ 
in regime A  
 in Fig.~\ref{figure03} together with the result of a fit of Eq.~\eqref{eq:s12fit}
to the DMRG data. In the shaded region, this fitting function describes the data very well.
We will use such fits to extract $\gamma_0$ as a function of $V,U$ and $\Gamma$.

Note that in a previous work by some of us \cite{busser13}, we demonstrated 
that an entangled state can be induced by sending a finite current through both dots
of a double quantum dot in a parallel geometry in the presence of a nonzero magnetic flux.
We also showed that this entangled state can be erased by decoupling one of the dots from
the environment, thus leading to a current flow through only the other dot. This results
in an exponential decay of spin correlations analogous to the case studied in the present work.

Finally, Fig.~\ref{figure02}(c) shows the concurrence as a function of time. 
Starting from a maximally  entangled state with $C_{12}(t=0)=1$, a transient regime similar to spin correlations with no dependence on $U$
is observed.
Then $C_{12}(t)$ rapidly decays and vanishes at a certain time $\tau_c$, whose value depends on $U$.
We use this time $\tau_c$ to define a decay rate $\gamma_c$ for the concurrence
\begin{equation}
\gamma_c = {\tau_c^{-1}}\,.
\end{equation}
The instantaneous drop of the concurrence to zero (as opposed to an
exponential, smooth decay) is known as sudden death of entanglement.
Having a  sharp or a smooth entanglement decay has been shown to depend on
the initial condition considered, on the type of noise, and on whether such
noise is applied locally or collectively to the two initially entangled
systems (see Ref.~\cite{ting09} for a review).

\begin{figure}[t]
\centerline{\includegraphics[width=8cm]{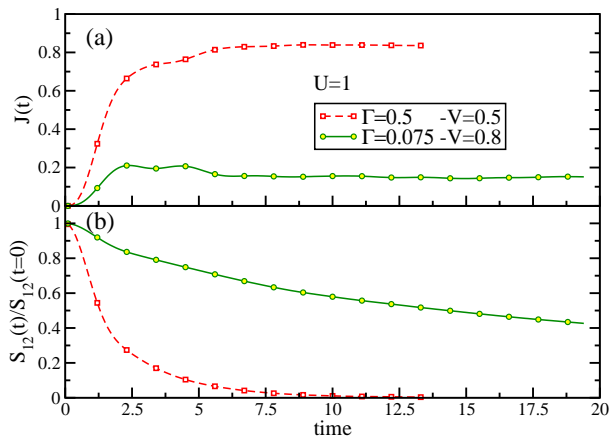}}
\caption{ (Color online)
This figure illustrates the effect of a finite and large $\Gamma$ in regime B ($V<U$)
by showing DMRG data for the two cases illustrated in Fig.~\ref{figure05}.
(a) Current versus time for $U=1$, $\Gamma=0.075, V=0.8$  
[solid lines with circles, corresponding to Fig.~\ref{figure05}~(a)] and $U=1, \Gamma=0.5, V=0.5$  
[dashed line with squares, corresponding to case Fig.~\ref{figure05}~(b)].  
(b) Spin correlations for the same parameters as in (a).
}\label{figure04a}
\end{figure}

\subsubsection{Effect of finite $\Gamma$}
\label{finitegamma}
At this point we would like to discuss the effect of the broadening of the resonant levels of QD1 due to the coupling to the environment,
or in other words, the consequences of a finite life-time for an electron to dwell on QD1.
Basically, the hybridization between QD1 and the reservoirs produces a finite level-width  given by 
\begin{equation}
\Delta(\omega) = \Gamma D(\omega)/D(E_F)\,. \label{broadening1}
\end{equation}
 This is the equilibrium broadening of a single resonant level and it is proportional to  $\Gamma$.
In our non-equilibrium problem, the levels are initially sharp and are expected to aquire
this width dynamically. This is precisely due to co-tunneling processes.
 These can, already in equilibrium, only be captured by a resummation, and
this physics is therefore beyond the regime of validity of the ME, which is second order in t'.

Two different situations can promote our system from regime B to A due to this broadening. 
The first one, represented in Fig.~\ref{figure05}(a), occurs for $U\gg \Gamma$ and  $V \lesssim U$ if
 the  broadened levels  overlap with the density of states of the leads. 
 The second case, in which the tunneling starts to be possible, occurs when $V<U$ but $\Gamma \sim U$, i.e., in the 
mixed-valence regime.
 In this situation, depicted in Fig.~\ref{figure05}(b), the tails of the broadened levels produce  tunneling 
processes for practically all $V>0$.
Numerical examples from DMRG simulations for these two situations are shown in Fig.~\ref{figure04a}.
We show both the spin correlations [Fig.~\ref{figure04a}(b)] and the current [Fig.~\ref{figure04a}(a)].
Note that, indeed, the current is finite in these examples and the spin correlations decay exponentially.

\subsection{Comparison of DMRG and master-equation results}
\label{sec:compare}
In this section, we compare results from DMRG and the master equation and we study the dependence of the 
rates $\gamma_0$ and $\gamma_c$ on model parameters.

\subsubsection{Time dependence of spin correlations in regimes A and B}

\begin{figure}[t]
\centerline{\includegraphics[width=7.5cm]{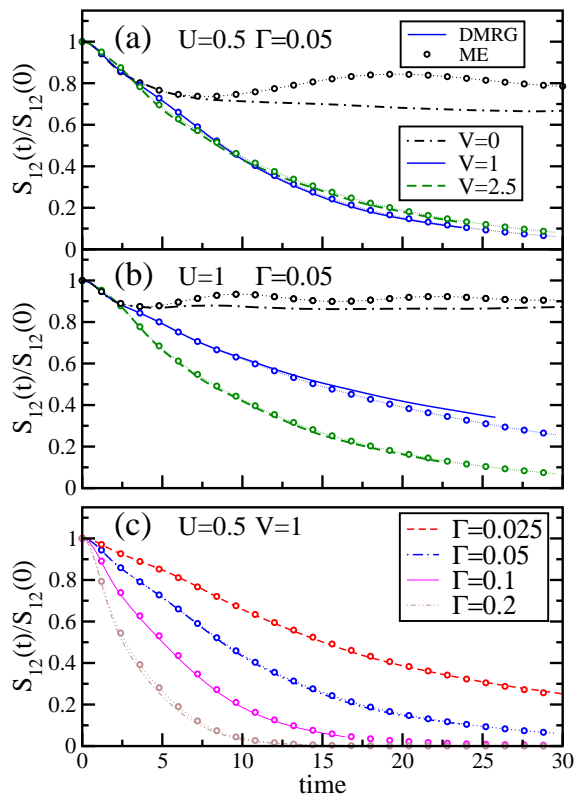}}
\caption{ (Color online) Comparison
between DMRG (solid lines) and master equation (short ME, symbols with dotted lines) results for the time dependence of
spin correlations $S_{12}(t)/S_{12}(0)$
for (a), (b) different voltages $V=0,1,2.5$ and (a) $U=0.5$, $\Gamma=0.05$ (b) $U=1$, $\Gamma=0.05$; (c) various values of $\Gamma=0.025,0.05,0.1,0.2$ ($N=35$).
}\label{figure04}
\end{figure}

In Fig.~\ref{figure04}, we present the spin correlations as a function of time with  DMRG results displayed as lines and ME results from the numerical solution of Eq.~\eqref{master1} as circles.
We show data for various voltages for 
$U/\Gamma=10$ and $U/\Gamma=20$ in Figs.~\ref{figure04}(a) and (b), respectively, and in Fig.~\ref{figure04}(c), we consider a fixed voltage $V=1$ and $U=0.5$, but different values of $\Gamma$.
The overall agreement between the DMRG and ME results is excellent in regime A. Small deviations are visible for 
large $\Gamma =0.2$ in Fig.~\ref{figure04}(c) or for $U=V$ in Fig.~\ref{figure04}(b) which is not surprising since 
on the one hand, the ME takes into account second-order processes in $t'$ only and on the other hand, $U=V$ is at the boundary
of regime A.

As can be seen from the plots, at initial times $t \lesssim 2$, a universal regime exists, in which basically all curves coincide for a given $\Gamma$. 
This  is because at small time scales the only relevant energy scale is the coupling of QD1 with the first site of each chain, which is given by $t'$. In addition, even if the curves   decays exponentially at later times, at times less or comparable to the environment correlation time $1/t_0\sim 1$,
they all exhibit a non-exponential decay typical of non-Markovian interactions. This initial deviation from a strictly exponential decay is captured by both the DMRG and 
the ME results. In the latter case, this is so because the ME Eq.~\eqref{master1} used here is more accurate than the Markovian ME, which would just predict an exponential decay from the very beginning [compare Eq.~\eqref{eq:s12fit}]. 

In the extreme case of $V=0$ in regime B, a qualitatively different behavior emerges since the spin correlations only
undergo a partial decay and then saturate at non-zero values  on the attainable time scales. Here, the deviations
between the DMRG and ME are quantitatively the largest. This is also the parameter regime, in which Kondo correlations
could become relevant on long times. This limit will be discussed in more detail in Sec.~\ref{sec:zero}.

\subsubsection{Decoherence rates}
\label{sec:rates}

We now turn to the analysis of the decay rates of the  spin correlations $S_{12}(t)$ and of the concurrence $C_{12}(t)$, extracted from the exponential decay
of the spin correlations or given by  the time at which $C_{12}(t)$ vanishes (compare Sec.~\ref{sec:expo}). 

\begin{figure}[t]
\centerline{\includegraphics[width=7.5cm]{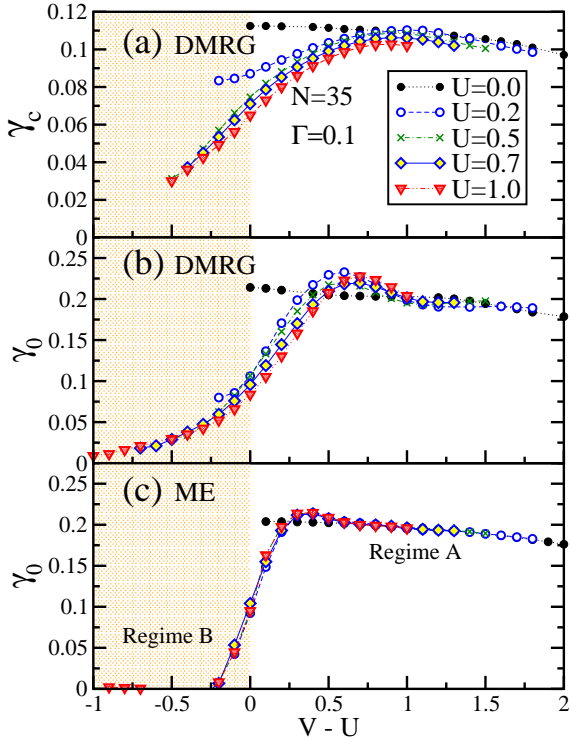}}
\caption{ (Color online)
Decoherence rates for spin correlations ($\gamma_0$) and concurrence ($\gamma_c$) for a fixed $\Gamma=0.1$ versus $V-U$ for different values of $U$
($U=0,0.2,0.5,0.7,1.0$).
(a) $\gamma_c$, DMRG data, (b) $\gamma_0$, DMRG data, (c)  $\gamma_0$, ME results.
 Plotting the rates versus $V-U$ results in a data collapse that works particularly well for the ME data.
Regime B is indicated by the shaded area.
} \label{figure06}
\end{figure}

\begin{figure}[t]
\centerline{\includegraphics[width=7.5cm]{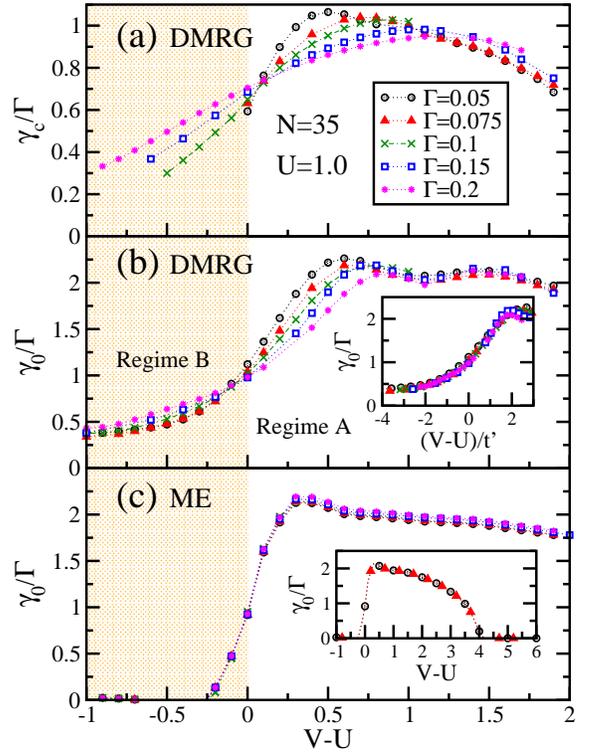}}
\caption{ (Color online)
Decoherence rates for spin correlations ($\gamma_0$) and concurrence ($\gamma_c$) for a fixed $U$ versus $V-U$ for different values of $\Gamma$
($\Gamma=0.05,0.075,0.1,0.15,0.2$).
(a) $\gamma_c/\Gamma$, DMRG data, (b) $\gamma_0/\Gamma$, DMRG data, (c)  $\gamma_0/\Gamma$, master equation (ME) results.
In regime A, we expect,  $\gamma_0 \propto 2\Gamma$ from Eq.~\eqref{eq:rate-regA}, which is confirmed by
the data shown here.
For large voltages, the semi-elliptical density of states of the tight-binding leads is resolved, resulting
in a decrease of $\gamma_0$ towards zero, which occurs at $V=4$ [see the inset in (c)].
Inset in (b): In the intermediate parameter range $V\sim U$ separating regimes A and B, a data collapse can be achieved by plotting the rate $\gamma_0/\Gamma$ versus $(V-U)/t'$.
Regime B is indicated by the shaded area. 
} \label{figure07}
\end{figure}

In Figs.~\ref{figure06} and~\ref{figure07}, we show these 
decay rates as a function of the applied bias. 
The regime B, in which 
the rate equation would predict a strictly vanishing $\gamma_0$, is indicated by a shaded area in these figures.

First, for a fixed coupling $\Gamma=0.1$,  Fig.~\ref{figure06} shows the results for the rates versus $V$ for various values of 
$U$. $\gamma_0$ is displayed in Fig.~\ref{figure06}(a), 
while Figs.~\ref{figure06}(b) and (c) contain results for $\gamma_c$ from DMRG and ME, respectively.
By plotting the data versus $V-U$, we can resolve the two regimes. We further observe that this simple rescaling 
results in a data collapse in regime A, which is particularly good for the ME data.
The data for $U=0$ and small $U=0.2$ show the strongest deviations.
When comparing Figs.~\ref{figure06}(b) and (c), we find that ME and DMRG are in good overall 
qualitative agreement, and in quantitative agreement in regime A for sufficiently large $V$.
The DMRG results for $\gamma_0$ in regime B are typically larger than the ones from the ME, which we interpret as
an indication that DMRG correctly accounts for virtual processes beyond the second order  in $t'$ for which the ME was constructed.
These include co-tunneling, not captured by the ME.

Regarding the rates for the concurrence, it is important to reiterate that these simply are  given by the inverse 
of the time $\tau_c$ beyond  which $C_{12}(t>\tau_c)=0$. Therefore, as an important result of our analysis, beyond this
time, the entanglement between the electrons in the  quantum dots has been erased.
This aspect is very robust against finite-size effects because of the instantaneous drop to zero of $C_{12}$.

In Fig.~\ref{figure07} we present the same quantities as in Fig.~\ref{figure06}, but now considering a fixed Coulomb repulsion $U=1$ and different values of $\Gamma$. 
As expected from the discussion of Sec.~\ref{sec:rate_eq}, both $\gamma_c$ and $\gamma_0$ are proportional to 
$\Gamma$ and hence we plot the rates as $\gamma_0/\Gamma$ and $\gamma_c/\Gamma$.
Equation~\eqref{eq:rate-regA} in fact predicts that in regime A, $\gamma_c/\Gamma\leq 2$. The decrease that occurs as $V$
increases is due to the change in the density of states of the leads and their finite band width [compare Eq.~\eqref{DOS}]. This is illustrated 
in the inset of Fig.~\ref{figure07}(c). 

Using the rescaling $\gamma_0/\Gamma$ and $V-U$, there remains a significant $\Gamma$ dependence in the
crossover region $V\sim U$, which is  evident in the DMRG shown in Fig.~\ref{figure07}(b). 
While we do not understand at present the origin, we observe that a universal behavior in the intermediate regime 
emerges if we plot the data versus $(V-U)/t'$. 

To summarize, the overall qualitative dependence of $\gamma_0$ on $U,V$ and $\Gamma$ can indeed be understood from the 
rate equation and Eq.~\eqref{eq:rate-regA}. The finite broadening results in smearing out the strict separation of 
regimes A and B. DMRG and ME results are in good agreement.

\subsection{Zero bias and the formation of Kondo correlations}
\label{sec:zero}
\begin{figure}[t]
\centerline{\includegraphics[width=8cm]{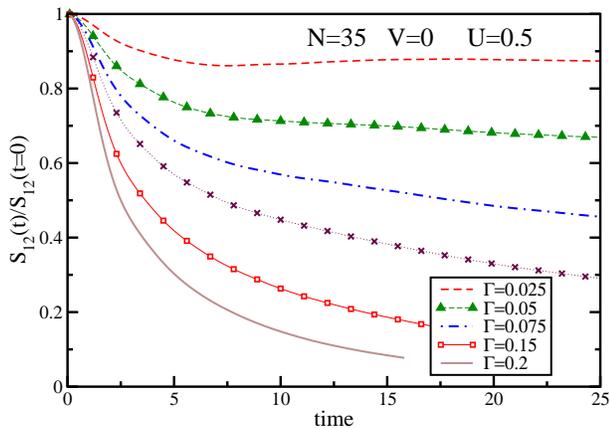}}
\caption{ (Color online) Spin correlations $S_{12}(t)$ versus time  
at zero voltage $V=0$ for $U=0.5$ and $\Gamma=0.025, 0.05,0.075,0.15, 0.2$ (top to bottom).
}
\label{figure11}
\end{figure}

We conclude the analysis by considering the special case of $V=0$. For this case, we do not observe a decay 
of spin correlations to zero on attainable time scales in our DMRG simulations for $U\gg \Gamma$ (compare Fig.~\ref{figure04}).
To further illustrate this point, we show $S_{12}(t)$ versus time for several values of $U/\Gamma$ at $V=0$
in Fig.~\ref{figure11}. Evidently, for $U\gg \Gamma$, no significant decay of the initial correlations takes place,
while as we approach the mixed-valence regime, we observe that $S_{12}(t)$  approaches zero already
on the short times and small systems studied here. The reason is that in the mixed-valence regime,  $T_K$ becomes comparable to $U$ and $\Gamma$ such that there
is no separation of time scales $1/\Gamma$ and $1/T_K$.
To get a handle on the  exponentially long time scale given by $\tau_K\propto 1/T_K$ 
and its dependence on $U/\Gamma$ we can either use
Haldane's expression for $T_K$ \cite{haldane78}
or refer to numerical results (see Ref.~\cite{dasilva08} for examples).
For the parameters of Fig.~\ref{figure04}(a) ($U=0.5$, $\Gamma=0.05$), such an estimate results in $\tau_K \gg 300$, respectively.
 This physics can be captured neither by the second-order weak-coupling expansion
employed in the ME nor with DMRG on tight-binding leads of a finite length for $U\gg \Gamma$.

Since the non-equilibrium dynamics at $V=0$ is driven by a local quench, namely
setting $t'$ to a non-zero value, it is reasonable to expect that for sufficiently long times,
ground-state correlations will  emerge. This implies that spin correlations between the electron in QD1
and electrons in the leads will eventually form, which results in the screening of the dot's magnetic moment.
This is a hallmark feature of Kondo physics.
These quasi-long range spin correlations are usually referred to as the Kondo cloud associated with a length
scale $\xi_K\sim 1/T_K$ \cite{sorensen96,sorensen05,holzner09,busser10,ghosh13}.
As a consequence of that, we expect the correlations between QD1 and QD2 to vanish in the long time limit.

\begin{figure}[t]
\centerline{\includegraphics[width=8cm]{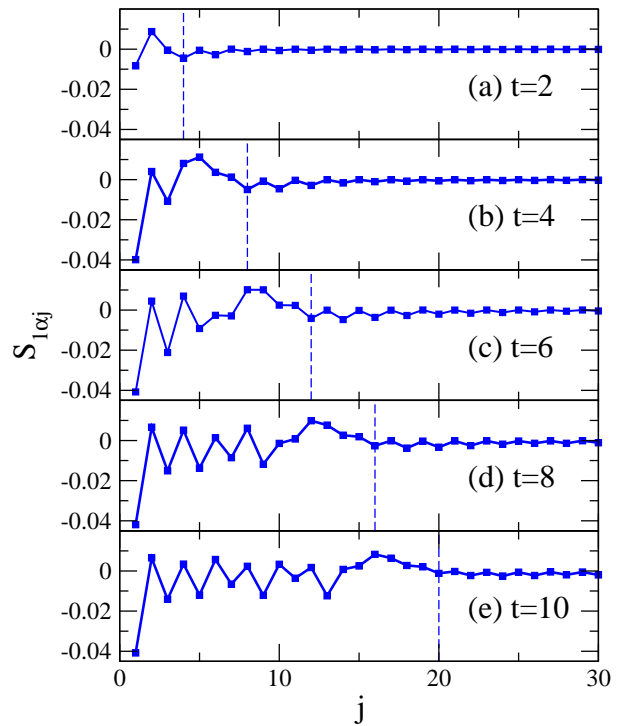}}
\caption{ (Color online) Spin correlations $S_{1\alpha j}$ between dot 1 
and sites $j$  in the leads as a function of position $j$ for $U=0.5$, $\Gamma=0.1$ and $V=0$ [for times $t\,t_0 =2,4,6,8,10$ shown in panels (a)-(e)]. The vertical lines indicate $j=2 t_0 t$ and the figure thus unveils 
 a typical light-cone structure that propagates through the leads at their Fermi velocity $v_F=2t_0$.}
\label{figure10}
\end{figure}

The dynamical emergence of such correlations is a very timely problem \cite{heyl,anders14,nuss14}. Here we
present DMRG results for the time dependence of spin correlations
between QD1 and the spin on sites $j$ in the lead $\alpha$
\begin{equation}
S_{1\alpha j}(t) = \langle \Psi(t) | \vec{S}_1 \cdot \vec{S}_{\alpha j} | \Psi(t) \rangle
\end{equation}
for $U=0.5$ and $\Gamma=0.1$.
The results shown in Fig.~\ref{figure10} unveil a typical light-cone structure \cite{calabrese07,heyl}: Correlations
become finite once the fastest excitations of the leads, which propagate with $v_F=2t_0$, have passed
a given site $j$. Such light-cones are commonly known in local and global quantum quenches \cite{heyl,manmana09,laeuchli08}
and have even been observed experimentally \cite{cheneau12}. Interestingly, there are additional emergent branches and oscillations.
Since at $V=0$, connecting QD1 to the leads is a local quantum quench, the quench energy is intensive,
and hence we expect to observe ground-state correlations (as computed in Refs.~\cite{holzner09,anders14})
to emerge at long times and sufficiently large sytems. This also explains the observation of large and negative
correlators close to the impurity.

In the regime of voltages $V< T_K$ and $U\gg \Gamma$, we expect that the current is initially zero
and then decreases at time scales $t\sim 1/T_K$.
Unfortunately,
finite-size effects are significant using a tight-binding  lead representation of the leads \cite{hm09} and therefore, we did not further pursue a quantitative comparison
with ground-state correlations and an analysis of transient currents in the low voltage regime, yet leave this for future research (see, e.g., the recent work by Nuss et al.~\cite{nuss14}), using either DMRG with Wilson leads \cite{dasilva08}
or the time-dependent numerical renormalization group method \cite{anders05,anders06}.

\section{Conclusions}
\label{conclusions}
In this work we studied the decay of a singlet state and thus a maximally entangled state
defined in a double-dot structure. The decoherence is induced by coupling one of the dots
to metallic leads, which may also be subject to a voltage difference. 
As a result of our combined analytical and numerical study, we identified
two qualitatively regimes. For $U \sim \Gamma$ and $V>0$, i.e., the mixed-valence regime,
either the effective levels on the quantum dot are resonant with the leads or the level
broadening gives rise to  tunneling processes. For $U \gg \Gamma$, direct tunneling
becomes possible only for $V\gtrsim U$, while for $V\lesssim U$, virtual tunneling
processes dominate on the accessible time scales.
Either type of process can cause spin flips on the dot coupled to leads, which are responsible 
for degrading the original spin correlations.

In the direct tunneling regime, we observe an exponentially fast decay of spin 
correlations. We presented an extensive analysis of the associated decay rate
on onsite interactions, tunneling rate, and voltage. These dependencies
can be qualitatively understood using a rate equation. In the small-voltage regime,
the decay of correlations is slower and in some cases (small $V$), only a partial 
loss of coherence occurs on the accessible time scales.

The results from the weak-coupling master equation approach and our numerical DMRG
simulations are in excellent quantitative agreement in the direct tunneling regime.

As a qualitative result, large voltages lead to an exponentially fast decay of the initial singlet,
whereas in the low-voltage regime, the singlet is much more stable and will presumably
only decay on time scales set by inverse Kondo temperature. This means that
one can use a large voltage to deliberately erase entanglement. On the other hand,
in a system with $U \gg \Gamma$  and  voltages smaller than Kondo temperature, 
likely the coupling to externel degrees of freedom (nuclear spin, etc.) will dominate the decay of the
singlet.

{\it Acknowledgment:} 
This project originated from discussions with  C. Neuenhahn and we are indebted to 
him for his important contributions in early stages of the project.
We further thank G. Burkard, S. Kehrein, F. Marquardt, L. Vidmar, and M. Zarea for helpful discussions,
and we are indebted to D. Schuricht for a critical reading  of the manuscript.
F.H.-M. thanks the Institute for Theoretical Physics II at the University of Erlangen for its hospitality, 
where parts of this work were performed. 
This work was supported by the Deutsche Forschungsgemeinschaft (DFG) through FOR~912 under grant-no.~HE5242/2-2 and through FOR~801, 
the Nanosystems Initiative Munich (NIM) (project No. 862050-2) and partially by the Spanish MICINN (Grant No. FIS2010-19998).
This work was also supported in part by National Science Foundation Grant No. PHYS-1066293 and the hospitality of the Aspen Center for Physics.

\appendix
\section{Master equation up to second order in the perturbative parameter}
The von-Neumann equation for the density operator of the total system $H=H_{\rm dots} +H_{\rm B} +H_{\rm hy} $ in the interaction picture (indicated by the superscript I), $\rho^{I}(t)$, reads as follows
\begin{equation}
\frac{d\rho^{I}(t)}{dt}=\frac{1}{i}[H_{\rm hy}(t),\rho^{I}(t)],
\label{total1}
\end{equation}
where $H_{\rm hy}(t) ={\mathcal U}^{-1}_{0}(t,t_0)H_{\rm hy} {\mathcal U}_{0}(t,t_0)$, $\rho^{I}={\mathcal U}^{-1}_{0}(t,t_0)\rho(t){\mathcal U}_{0}(t,t_0)$ and the free evolution operator is ${\mathcal U}_{0}(t,t_0)=\exp[-iH_{0}(t-t_0)]$ with $H_0=H_{\rm dots} +H_{\rm B}$. To simplify the notation, we set $\rho^{I}(t)=\rho(t)$. We can integrate Eq.~(\ref{total1}) between $t_0$ and $t$, with $t-t_0 =\Delta t$. After some iterations and taking the trace over the bath's degrees of freedom, this leads to the following equation
\begin{eqnarray}
&&\Delta \rho_{\rm dots} (t)=\frac{1}{i}\int_{t_0}^{t}d\tau \mbox{Tr}_{\rm B}\{[ H_{\rm hy}(\tau),\rho(t_0)]\}+ {\left( \frac{1}{i} \right)}^2 \cr
&\times&\int_{t_0}^{t}d\tau\int_{t_0}^{\tau} d\tau' \mbox{Tr}_{\rm B}\{[{H}_{\rm hy}(\tau),[{H}_{\rm hy}(\tau'),\rho(\tau')]]\},
\label{total2}
\end{eqnarray}
where $\rho_{\rm dots} (t)=\mbox{Tr}_{\rm B}\{\rho(t)\}$ is the reduced density operator of the system $H_{\rm dots}$ and
\begin{equation}
\Delta \rho_{\rm dots} (t)=\rho_{\rm dots} (t)-\rho_{\rm dots} (t_0).
\end{equation}
Equation~(\ref{total2}) is exact, but some assumptions have to be made in order to express it in a more simple way.  

Choosing an initially decorrelated condition between the system and the environment, $\rho(t_0)=\rho_{\rm dots} (t_0)\otimes \rho_{\rm B} (t_0)$, and considering that the average value in $\rho_{\rm B} (t_0)$ of the perturbation term $H_{\rm hy}(t)$ is zero, 
\begin{equation}
\mbox{Tr}_{\rm B}[{H}_{\rm hy} (t_0)\rho_{\rm B}(t_0)]=0,
\end{equation}  
so that the first term in Eq.~(\ref{total2}) can be eliminated.
After the change of variable ${\mathcal T}=\tau$ and $s=\tau-\tau'$, Eq.~(\ref{total2}) becomes, 
\begin{eqnarray}
\rho_{\rm dots}(t)&=&\rho_{\rm dots} (t_0)-\int_{t_0}^{t}d{\mathcal T}\int_{0}^{{\mathcal T}-t_0} d\tau \mbox{Tr}_{B}\{[{H}_{\rm hy} ({\mathcal T}),\cr 
&&
[{H}_{\rm hy}({\mathcal T}-\tau) ,\rho({\mathcal T}-\tau)]]\}.
\label{total3chap3}
\end{eqnarray}
The evolution equation for the reduced density operator can be obtained by taking a derivative with respect to $t$ in  Eq.~(\ref{total3chap3}) 
\begin{eqnarray}
&& \frac{d\rho_{\rm dots} (t)}{dt}=   \label{total4}\\ 
&&-  \int_{0}^{t-t_0}d\tau \mbox{Tr}_{\rm B}\bigg([{H}_{\rm hy }(t),[{H}_{\rm hy} (t-\tau),\rho_{}(t-\tau)]]\bigg),
\nonumber
\end{eqnarray}
with the initial condition given by $\rho_{\rm dots} (t_0)$. In order to transform Eq.~(\ref{total3chap3}) into an equation for $\rho_{\rm dots}$ that is local in time, it is necessary to perform a \textit{Markovian approximation} over the time evolution of the system. In this approximation, the evolution of $\rho$ from $t_0$ to $t$ is neglected, provided that the domain of integration time $\Delta t=t-t_{0}$ is small enough in comparison with the evolution time scale of the system $T_{A}$ ($\Delta t \ll T_{A}$), where $T_A$ is the relaxation time scale of the system, of the order of $1/\Gamma$. Notice that this Markovian approximation, which is related to the evolution time scale of the density operator, is not the same as the Markovian approximation over the bath evolution time scale. In the latter, the correlation time of the bath, $\tau_{c}$, is assumed to be  much smaller  than characteristic   time scales of the system ($\tau_{c}\ll T_{A}$). In this derivation  the Markovian approximation is considered over the density operator, but not over the bath. In that way, the density operator appearing on the right hand side of Eq.~(\ref{total4}) is already local in time, but it is still composed of three terms:
\begin{equation}
\rho(t)=\rho_{\rm dots} (t)\otimes \mbox{Tr}_{\rm dots} \{\rho(t)\}+\rho_{\rm correl}(t).
\end{equation} 
The term $\rho_{\rm correl}(t)$, which describes the correlation between the system and the bath at time $t$, can be neglected with the assumption that $\tau_{C}\ll \Delta t$, considering that the correlations at time $t$ disappear after a time which is approximately equal to $\tau_c$. This is the so-called Born approximation, which is only valid up to order $\mathcal{O}(t'^2)$ in the perturbation parameter $t'$ \cite{atomphotoninteractions,breuer1999}. 

With these approximations and choosing $t_0 =0$, Eq.~(\ref{total4}) becomes just Eq.~(\ref{total5}) shown in the main text. 

\section{Fermi's Golden Rule and master equation in the Markov limit}
\label{sec:markov}

Let us now consider Eqs.~(\ref{correlplus}) and (\ref{correlminus}) in the continuum limit
\begin{eqnarray}
&&G_\alpha^- (t-\tau)={t'}^2\sum_k 
(1-f_\alpha(\epsilon_k))e^{-i\epsilon_k (t-\tau)}\cr
&=&{t'}^2\int_0^\infty d\omega D(\omega)(1-f_\alpha(\omega))e^{-i\omega(t-\tau)},
\label{correlplus2}
\end{eqnarray}
and
\begin{eqnarray}
&&G_\alpha^+ (t-\tau)={t'}^2\sum_k f_\alpha(\epsilon_k)e^{i\epsilon_k (t-\tau)}\cr
&=&{t'}^2\int_0^\infty d\omega D(\omega)f_\alpha(\omega)e^{i\omega(t-\tau)}\,.
\label{correlminus2}
\end{eqnarray}
Inserting these expressions in Eq.~(\ref{rates1}), and extending the limits of the time integrals to infinity we get the following expressions for the rates 
\bea
\gamma^{-}_{\alpha\sigma}&=&{t'}^2\int_0^\infty d\omega D(\omega)(1-f_\alpha(\omega)) \bigg(\delta(\omega-\Omega_{\sigma})\cr
&+&\frac{i}{2\pi}{\mathcal P}\frac{1}{\omega-\Omega_{\sigma}}\bigg)=\Gamma^-_{\alpha\sigma}+i\hat{\Gamma}^{-}_{\alpha\sigma};\cr
\gamma^{+}_{\alpha\sigma}&=&{t'}^2\int_0^\infty d\omega D(\omega)f_\alpha(\omega) \bigg(\delta(\omega-\Omega_{\sigma})\cr
&-&\frac{i}{2\pi}{\mathcal P}\frac{1}{\omega-\Omega_{\sigma}}\bigg)=\Gamma^+_{\alpha\sigma}+i\hat{\Gamma}^{+}_{\alpha\sigma},
\label{rates2}
\eea
where we have considered that $\int_0^\infty d\tau e^{i\omega\tau}=\delta(\omega)+\frac{i}{2\pi}{\mathcal P}\bigg(\frac{1}{\omega}\bigg)$. Taking this into account, new rates $\Gamma_{\alpha\sigma}^{\pm }$ can effectively be defined. The real parts $\Gamma_{\alpha\sigma}^{\pm}$ are given in Eq.~(\ref{rates3}). 

As mentioned above, the extension to infinity of the time limits in Eq.~(\ref{rates1}) is justified when the decay of the functions $G_\alpha^{+}(t-\tau)$ and $G_\alpha^{-}(t-\tau)$ is very fast compared to the time evolution of the quantum dot system, approximately given by $1/{t'}^2$.  In this limit, the coefficients given in Eq.~(\ref{rates1}) rapidly converge to a constant value given by Eq.~(\ref{rates2}). 
The first part of Eq.~(\ref{rates2}), corresponding to a real quantity, corresponds to the system's dissipation rates calculated through  Fermi's Golden Rule, and constitutes the relevant contribution to the decoherence process. The second term of Eq.~(\ref{rates2}) gives rise to a contribution that, once inserted in Eq.~(\ref{master3}), gives rise to terms that can be recast in the form of a coherent evolution of the system's degrees of freedom with an effective Hamiltonian given by 
\bea
H_\eff=\sum_{\alpha\sigma}\hat{\Gamma}^{+}_{\alpha\sigma}d_{1\sigma}^\dagger d_{1\sigma}-\sum_{\alpha\sigma}\hat{\Gamma}^{-}_{\alpha\sigma}d_{1\sigma}^\dagger d_{1\sigma}.
\eea
Rewriting the master equation Eq.~(\ref{master2}), we find 
\begin{eqnarray}
&&\frac{d\rho_{\rm dots}  (t)}{dt}=  
-i[\tilde H_{\rm dots} ,\rho_{\rm dots} (t) ]\cr
&&+\sum_{\alpha,\sigma}\Gamma^{+}_{\alpha\sigma}
\bigg( [d_{1\sigma}^\dagger ,\rho_{\rm dots} (t)  d_{1\sigma}]
+[d_{1\sigma}^\dagger \rho_{\rm dots} (t) ,d_{1\sigma}]\bigg)\cr 
&&+\sum_{\alpha,\sigma}\Gamma^{-}_{\alpha\sigma} \bigg(
[d_{1\sigma}\rho_{\rm dots} (t) ,d_{1\sigma}^{\dagger}]
+[d_{1\sigma},\rho_{\rm dots} (t) d_{1\sigma}^{\dagger}]\bigg)\cr 
&&+{\mathcal O}({t'}^4),
\label{master3}
\end{eqnarray}
where $\tilde{H}_{\rm dots}=H_{\rm dots}+H_\eff$. 

To proceed further, the master equation Eq.~(\ref{master3}) should be expressed in terms of the system's (i.e., the  quantum dots) unperturbed eigenbasis, spanned  by the following four eigenvectors: $\{|0\rangle,|\uparrow\rangle,|\downarrow\rangle,|\uparrow,\downarrow\rangle\}$, which represent  states with zero, one spin up, one spin down, and two electrons in the quantum dot, respectively. To make less involved the notation, we shall relabel these four basis members as $|n\rangle$, for $n=0,\cdots,3$. In terms of these, we find that the master equation can be rewritten as 
\begin{eqnarray}
&&\frac{d\rho_{\rm dots}  (t)}{dt}=-i[\tilde{H}_{\rm dots},\rho_{\rm dots} (t) ] \cr 
&& +\sum_{\alpha,\sigma}\Gamma^{+}_{\alpha\sigma} \bigg(2D_{\sigma}-\rho_{\rm dots} B_{\sigma}
-B_{\sigma}\rho_{\rm dots}\bigg) \cr 
&& +\sum_{\alpha,\sigma}\Gamma^{-}_{\alpha\sigma} \bigg( 2E_{\sigma}- A_{\sigma}\rho_{\rm dots}-\rho_{\rm dots} A_{\sigma}\bigg),\nonumber
\label{master4}
\end{eqnarray}
where we have defined 
\bea
A_{\sigma}&=&d_{\sigma}^\dagger d_{\sigma}=\delta_{\sigma, 2}(|3\rangle\langle 3|+|2\rangle\langle 2|)+\delta_{\sigma, 1}(|3\rangle\langle 3|+|1\rangle\langle 1|)\cr
B_{\sigma}&=&d_{\sigma} d_{\sigma}^\dagger=\delta_{\sigma, 2}(|0\rangle\langle 0|+|1\rangle\langle 1|)+\delta_{\sigma, 1}(|0\rangle\langle 0|+|2\rangle\langle 2|)\cr
D_{\sigma}&=&d^\dagger_{\sigma}\rho_{\rm dots} d_\sigma=\delta_{\sigma, 1}(|1\rangle\langle 0|+|3\rl 2|)\rho_{\rm dots}(|0\rl 1|+|2\rl 3|)\cr
&+&\delta_{\sigma, 2}(|2\rangle\langle 0|+|3\rl 1|)\rho_{\rm dots}(|0\rl 2|+|1\rl 3|);\cr
E_{\sigma}&=&d_{\sigma}\rho_{\rm dots} d^\dagger_\sigma=\delta_{\sigma, 1}(|0\rangle\langle 1|+|2\rl 3|)\rho_{\rm dots}(|1\rl 0|+|3\rl 2|)\cr
&+&\delta_{\sigma, 2}(|0\rangle\langle 2|+|1\rl 3|)\rho_{\rm dots}(|2\rl 0|+|3\rl 1|)\,.
\eea
Thus, within the Markov approximation, the matrix elements of the reduced density matrix evolve as follows
\bea
\frac{d\rho_{00}}{dt}&=&-i\langle 0|[\tilde H_{\rm dots},\rho_{\rm dots}]|0\rangle+2\tilde{\Gamma}^-_{1}\rho_{11}\cr 
			&& +2\tilde{\Gamma}_{2}^-\rho_{22} - 2(\tilde{\Gamma}_{2}^++\tilde{\Gamma}_{1}^+)\rho_{00}\cr
\frac{d\rho_{11}}{dt}&=&-i\langle 1|[\tilde H_{\rm dots},\rho_{\rm dots}] |1\rangle\cr 
&& +\bigg(2\tilde{\Gamma}_{1}^+\rho_{00}-2(\tilde{\Gamma}_{1}^-+\tilde{\Gamma}_{2}^+)\rho_{11}
+2\tilde{\Gamma}^-_{2}\rho_{33}\bigg)\cr
\frac{d\rho_{22}}{dt}&=&-i\langle 2|[\tilde H_{\rm dots},\rho_{\rm dots}]|2\rangle \cr 
&&+\bigg(2\tilde{\Gamma}_{2}^+\rho_{00}-2(\tilde{\Gamma}_{1}^++\tilde{\Gamma}_{2}^-)\rho_{22} +2\tilde{\Gamma}^-_{1})\rho_{33}\bigg)\cr
\frac{d\rho_{33}}{dt}&=&-i\langle 1|[\tilde H_{\rm dots},\rho_{\rm dots}]|3\rangle+2\tilde{\Gamma}_{2}^+\rho_{11}+2\tilde{\Gamma}_{1}^+\rho_{22}\cr
&-&2(\tilde{\Gamma}^-_{2}+\tilde{\Gamma}^-_{1})\rho_{33}\bigg)\cr
\frac{d\rho_{12}}{dt}&=&-i\langle 1|[\tilde H_{\rm dots},\rho_{\rm dots}]|2\rangle-\sum_{\sigma}
(\tilde{\Gamma}^-_{\sigma}+\tilde{\Gamma}^+_{\sigma})\rho_{12}
\label{evol1}
\eea
Here,  we have considered that $\rho_{nm}=\Ttr\{\rho_{\rm dots}(t)|n\rangle\langle m|\}$, and $d^\dagger_{1\uparrow}=d^\dagger_{11}=|1\rangle\langle 0|+|3\rangle\langle 2|$,  $d_{1\uparrow}=d_{11}=|0\rangle\langle 1|+|2\rangle\langle 3|$, $d^\dagger_{1\downarrow}=d^\dagger_{12}=|2\rangle\langle 0|+|3\rangle\langle 1|$,  $d_{1\downarrow}=d_{12}=|0\rangle\langle 2|+|1\rangle\langle 3|$. 
Thus, $\rho_{nn'}$ is an element of the density matrix $\rho_{\rm dots}$ belonging to the Hilbert space of QD1, such density matrix corresponding to one of the matrices $\rho^{(1),n}(t)$, with $n=1,\cdots,4$, with initial states given by Eq.~(\ref{eq:four}). 
Also, we have defined $\tilde{\Gamma}_\sigma^\pm$ 
\bea
\tilde{\Gamma}_\sigma^\pm=\sum_\alpha \Gamma_{\alpha,\sigma}^\pm\,.
\label{eq:rate2}
\eea
It is interesting to notice that the element $\langle 1|[\tilde H_{\rm dots},\rho_{\rm dots}]|2\rangle=0$ during the evolution, such that the evolution equation for the coherences is just written as Eq.~(\ref{eq:rate}).


\bibliographystyle{apsrev4-1}
\bibliography{decoh-ref}
\end{document}